\documentclass[superscriptaddress,
twocolumn,
amsmath,amssymb]{revtex4}
\usepackage{epsfig}
\usepackage{bm}
\usepackage{times}
\usepackage[usenames]{color}
\begin{document}
%\setstretch{2}
\title{Hilbert transform based analyses on ship-rocking signals}

\author{Wei Huang} 
\affiliation{Department of Modern physics, University of Science and
  Technology of China, Hefei, Anhui 230026, China}

\author{Yu-jian Li}
\affiliation{ China Satellite Maritime Tracking and Control
  Department, Jiangyin 214431, China}

\author{Deyong Kang}
\affiliation{ China Satellite Maritime Tracking and Control
  Department, Jiangyin 214431, China}

\author{Zhi Chen} 
\affiliation{Department of Modern physics, University of Science and
  Technology of China, Hefei, Anhui 230026, China}
%\date{\today}

%\pacs{05.40.Ca, 75.40.Gb, 73.50.Td, 75.40.Mg}

\begin{abstract}
The ship-rocking is a crucial factor which affects the accuracy of the
ocean-based flight vehicle measurement. Here we have analyzed four
groups of ship-rocking time series in horizontal and vertical
directions utilizing a Hilbert based method from statistical
physics. Our method gives a way to construct an analytic signal on the
two-dimensional plane from a one-dimensional time series. The analytic
signal share the complete property of the original time series. From
the analytic signal of a time series, we have found some information
of the original time series which are often hidden from the view of
the conventional methods. The analytic signals of interest usually
evolve very smoothly on the complex plane. In addition, the phase of
the analytic signal is usually moves linearly in time. From the
auto-correlation and cross-correlation functions of the original
signals as well as the instantaneous amplitudes and phase increments
of the analytic signals we have found that the ship-rocking in
horizontal direction drives the ship-rocking in vertical direction
when the ship navigates freely. And when the ship keeps a fixed
navigation direction such relation disappears. Based on these results
we could predict certain amount of future values of the ship-rocking
time series based on the current and the previous values. Our
predictions are as accurate as the conventional methods from
stochastic processes and provide a much wider prediction time range.

\end{abstract}
\maketitle 

%\section*{Introduction}

In the ocean-based flight vehicle measurement, a very important task
is to characterize special properties of the objective from massive
quantity of data. Based on this one may have a way to increase the
accuracy of the measurement. Unfortunately, there are many factors in
the measurement which could affect the accuracy of the results. And it
is even worse since these factors may correlate with one and
another. As a result, the time series obtained in the measurements
often display the following characteristics: periodicity, may be
non-stationary, noisy and may contain certain stochastic components:
such as outliers which are hard to explain their origin and random
jumps in the data. To increase the accuracy of the measurement, much
effort has been done to study two crucial time series: the ship
rocking and the ship deformation. The current strategy is to make
appropriate corrections based on these two time series, and thus to
reduce or even eliminate the periodicity in the signals of
interest. However, after these corrections the measurement results
still show considerable difference from those basement standard
results obtained through other ways, such as GPS, laser, etc.

In this paper we consider the ship-rocking time series based on the
approach from statistical physics. To those time series which show
strong periodicity, an effective method is to investigate the
instantaneous analytic signal of a time series~\cite{ChenPRE06}. The
instantaneous analytic signal is obtained based on a Hilbert
transform. The original time series as well as its Hilbert transform
construct an analytic signal at each measurement time. On the complex
plane one can calculate the instantaneous amplitude and the
instantaneous phase of the analytic signal. Such instantaneous
amplitude and phase may reveal certain intrinsic properties of systems
which are not seen from the conventional methods. For example,
previous research shows that, the change of the correlation in human
auto-regulation may not happen in the original signal, however it does
show in the correlation of its instantaneous phase
increments~\cite{ChenPRE06}. Here, similarly we investigate the
characteristics of the analytic signal. We hypothesize that one may find
some properties which are hidden from the view of traditional methods.

\section{Data}

From our collaborators we have obtained four groups of ship-rocking
time series: GX1129，JL839，RW839，RW840. The data starting
with ``GX'' are those obtained when the ship navigates freely, while
others are those obtained when the ship navigates while maintaining a
fixed orientation. All other information is unknown to us. For each
group we have the time series in three directions of the Euler angles:
the navigation direction, the horizontal direction and the vertical
direction. In the following we mark the navigation direction as
``KC'', the horizontal direction as ``OC'', and the vertical direction
as ``PC''. If not specially indicated, the unit of all angles is in 
radian. All time series are measured every 50 milliseconds.

\begin{figure}
\epsfysize=0.9\columnwidth{\rotatebox{-90}{\epsfbox{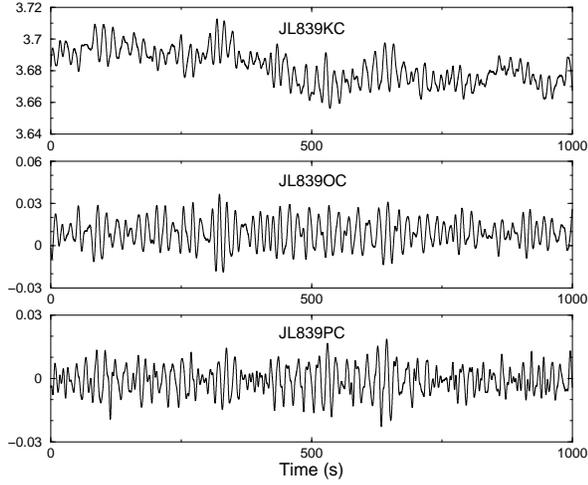}}}
\caption{The ship-rocking time series in three directions for the
  group JL839. }
\label{fig1}
\end{figure}

As shown in Fig.~\ref{fig1}, the most apparent characteristic of the
ship-rocking signal is periodicity. For all four groups of data
available, the period in signals varies from around 6 seconds to more
than 20 seconds. Such period also fluctuates at different measurement
position of the same time series. In the KC direction, the situation
becomes more complicated since the running average of the time series
also varies at different time positions, e.g., as shown in
Fig.~\ref{fig1}. Such kind of signal apparently shows nonstationarity
at different times. Thus many conventional methods may not apply. One
possible approaches to such kind of signals is to apply the Detrended
Fluctuation Analysis (DFA), which is very popular in recent years to
quantify characteristics in non-stationary
signals~\cite{PengPRE94,ChenPRE02,ChenPRE05}. Another approach is to
reduce such nonstationarity in the frequency domain, i.e., remove the
low frequency part in the Fourier transform of the signal, then apply
the reverse Fourier transform to obtain the new, stationary time
series. However, the validity of these approaches to the current data
has to be carefully examined before any convincing results could be
accepted. This will be done in the future paper. Here we instead focus
on stationary ship-rocking time series in two other directions: OC and
PC. We will investigate the original signals as well as their
instantaneous amplitudes and phases. We expect that certain intrinsic
properties of the current system may show in the auto-correlation
functions and the cross-correlation functions of these signal.

\section{ The basic technique}

Our results in this paper are based on the Hilbert transform (HT) to a
stationary time series. To any stationary time series $s(t)$, its HT
is defined as: 
\begin{eqnarray}
\tilde{s}(t) \equiv
\frac{1}{\pi}P\int_{-\infty}^{\infty}\frac{s(\tau)}{t-\tau}d\tau,
\label{eqn1}
\end{eqnarray}
where P denotes the Cauchy principal value. In the frequency domain,
the Fourier component at any frequency $f$ for the Fourier transform
of $\tilde{s}(t)$ can be obtained very easily from the Fourier
transform of $s(t)$, i.e., that of $s(t)$ at the same frequency $f$
rotates $90^{\circ}$ clockwise (for $f>0$) or anticlockwise (for
$f<0$) in the complex plane. As a simple example, if $s(t)=\sin(\alpha
t)$, then one would obtain $\tilde{s}(t)=\cos (\alpha t)$. Based on
this fact, one could define an ``analytic signal'' for this time
series: 
\begin{eqnarray}
S(t)\equiv s(t) + i \tilde{s}(t)=A(t)e^{i\varphi(t)},
\label{eqn2}
\end{eqnarray}
where $A(t)$ and $\varphi(t)$ are the instantaneous amplitude and
phase of $s(t)$, respectively.

\section{Characteristics of the analytic signals}

\subsection{the original signal and its Hilbert transform}\label{hilbert} 

\begin{figure}
\epsfysize=0.9\columnwidth{\rotatebox{-90}{\epsfbox{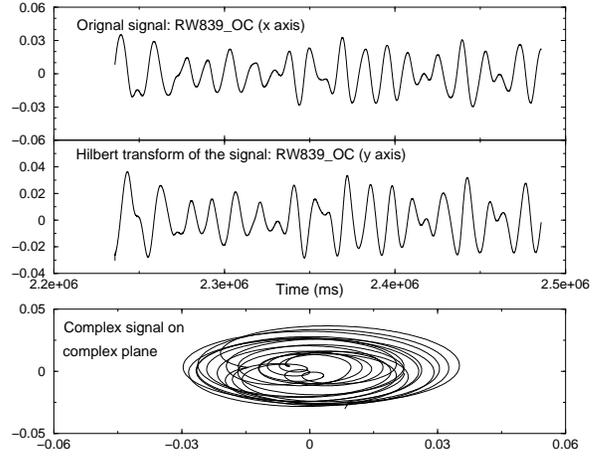}}}
\caption{The original ship-rocking time series and its Hilbert
  transform, as well as the corresponding analytic signal for the
  group RW839 in the horizontal (OC) direction. }
\label{fig2}
\end{figure}

Following the procedure presented in the last section, one would
obtain two additional time series from the original time series $s(t)$:
its amplitude and its phase. In Fig.~\ref{fig2} we show the patterns
of the ship-rocking time series for the group RW839 in the OC
direction. Similar behaviors are observed in signals of other groups
and in other directions. In the analyses, we always remove the zero
frequency part in the Fourier transform of $s(t)$ (which is related to
the average of $s(t)$). Thus in the complex plane, when the time
proceeds, the analytic signal will rotate around the origin of the
plane. At any time, the distance of the point on the analytic signal
to the origin is the instantaneous amplitude, and the angle of the
point to the positive direction of the x-axis is the phase.

From the lowest panel of Fig.~\ref{fig2} one could find that the
evolution of the analytic signal $S(t)$ is very slow when seen in the
complex plane. It usually follows the similar behavior when rotating
around the origin. (This can be further verified through the behavior
of the instantaneous phase shown in Sec.~\ref{amp_pha}.) However
$s(t)$ could evolve very slowly at certain times, e.g., at the though
and the peak of the wave. In contrast, $s(t)$ may also evolve very
rapidly at the middle part between the though and the peak of the
wave. Based on this characteristic, it may be possible to predict the
behavior of the ship-rocking time series in certain future times from
the data at the present and in previous times. We will show the
details in the Sec.~\ref{predict}.

\subsection{Behaviors of the instantaneous amplitude and phase}\label{amp_pha}

\begin{figure}
\epsfysize=0.49\columnwidth{{\epsfbox{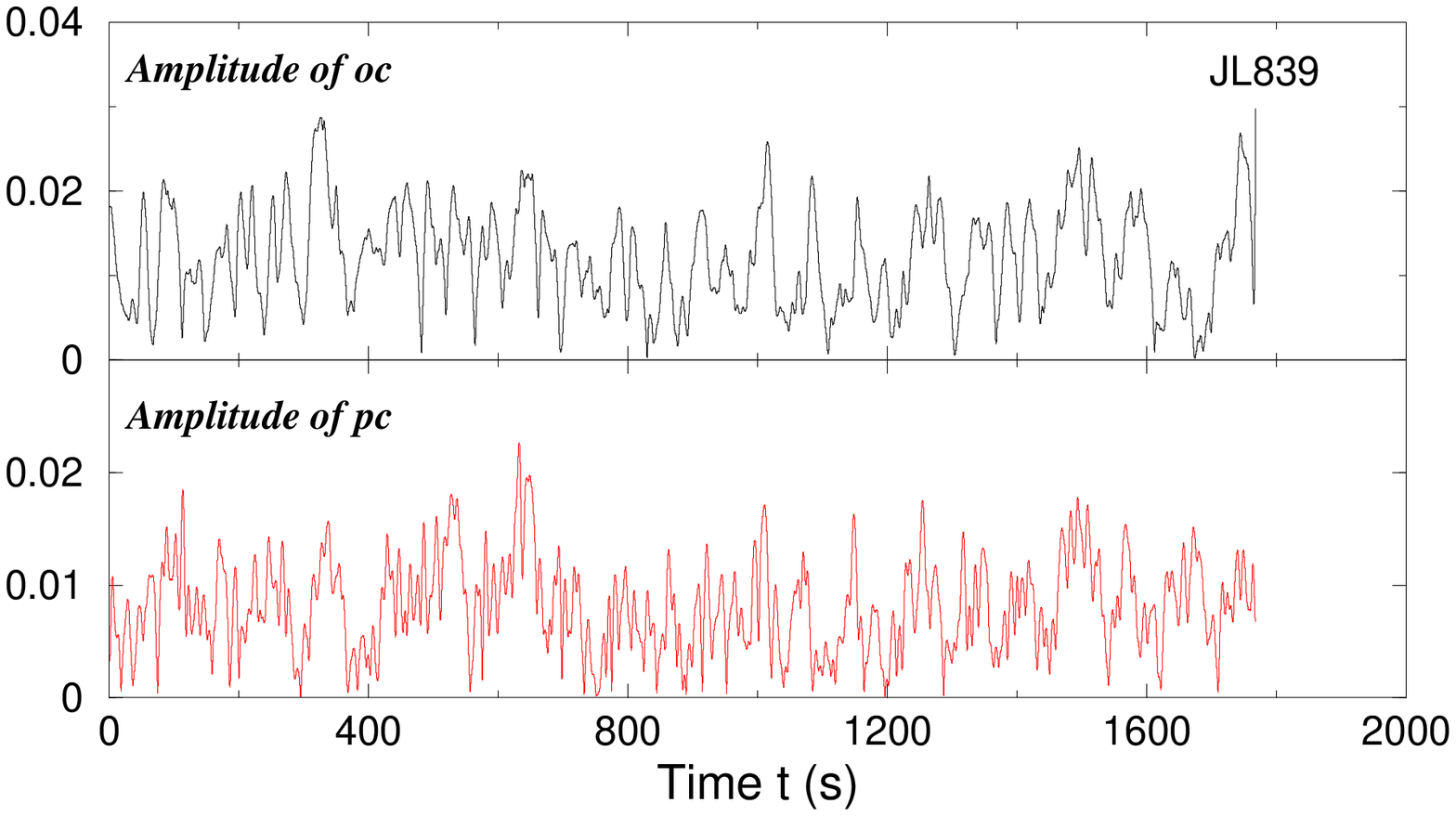}}}
\epsfysize=0.49\columnwidth{{\epsfbox{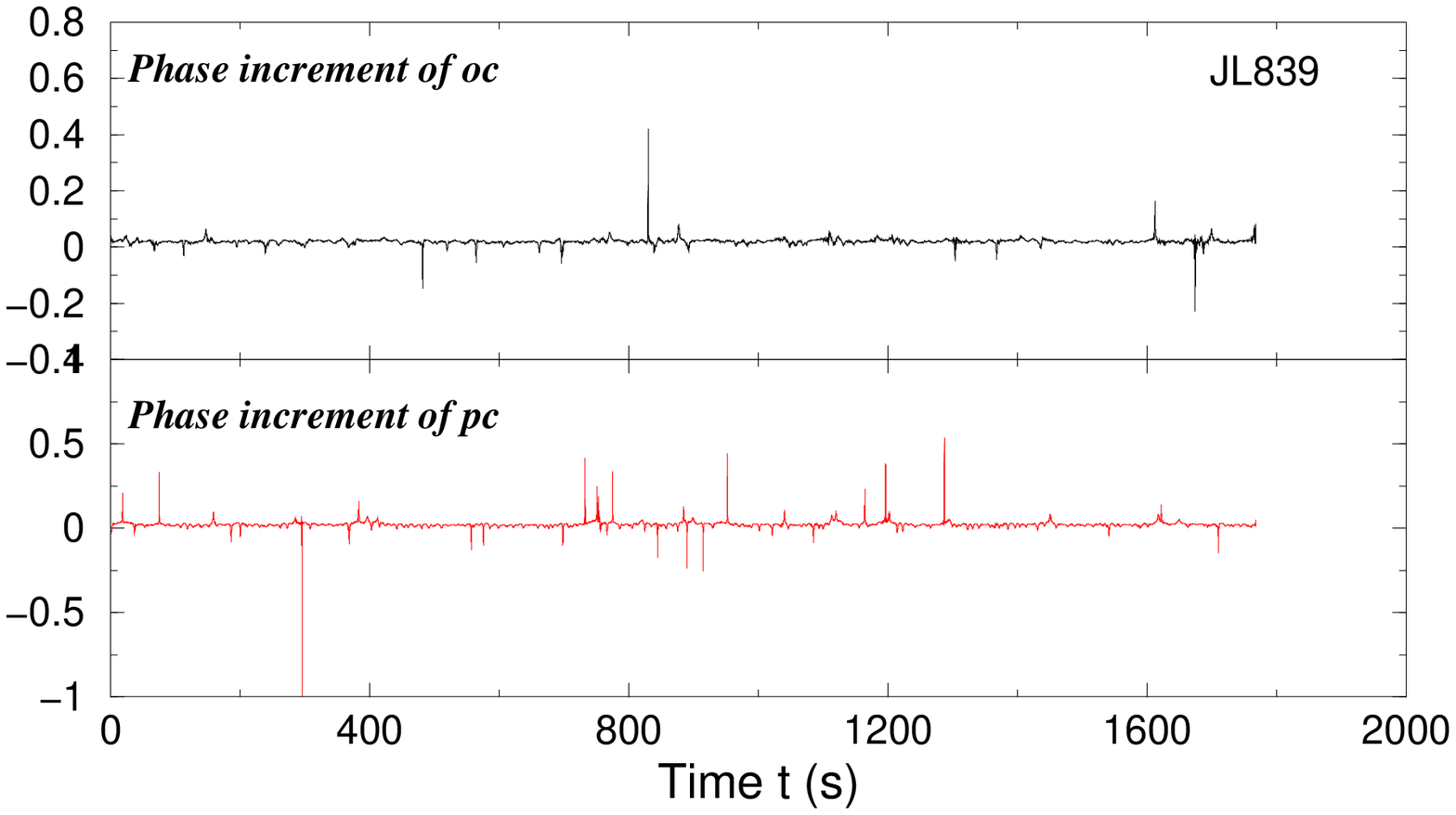}}}
\caption{The instantaneous amplitudes and phase increments for the
  group JL839 in OC and PC directions. }
\label{fig3}
\end{figure}

We have investigated the characteristics in the instantaneous
amplitudes and instantaneous phases in all four groups of the
ship-rocking time series. We find that they behave similarly between
different groups and in different directions. As an example, in
Fig.~\ref{fig3} we show how the instantaneous amplitudes and the phase
increments during unit time interval (50 ms) in OC and in PC
directions evolve in time for the group JL839.

As shown in Fig.~\ref{fig3}, the instantaneous amplitude of the
original data is non-negative and still displays periodicity in
time. However its shape is more irregular and its periods are much
larger than those of the original data. In contrast, the instantaneous
phase does not show any periodicity. In a large enough time window,
one could find that the instantaneous phase depends approximately
linearly on the time. Thus in Fig.~\ref{fig3} we show instead the
phase increments of the original data during the unit time
interval. In most times the phase increments are approximately a
constant. However, we also notice that at certain times there are some
sudden jumps in the value of phase increments. Such jumps could be
some outliers contributed by certain stochastic noise. However it
could also due to certain human or environmental activities on the
ship. We are working closely with our collaborators to investigate the
origin of these sudden jumps.

\subsection{Auto-correlation functions for the original data and their instantaneous amplitudes and phase increments}\label{autocorr}

\begin{figure}
\epsfysize=0.8\columnwidth{{\epsfbox{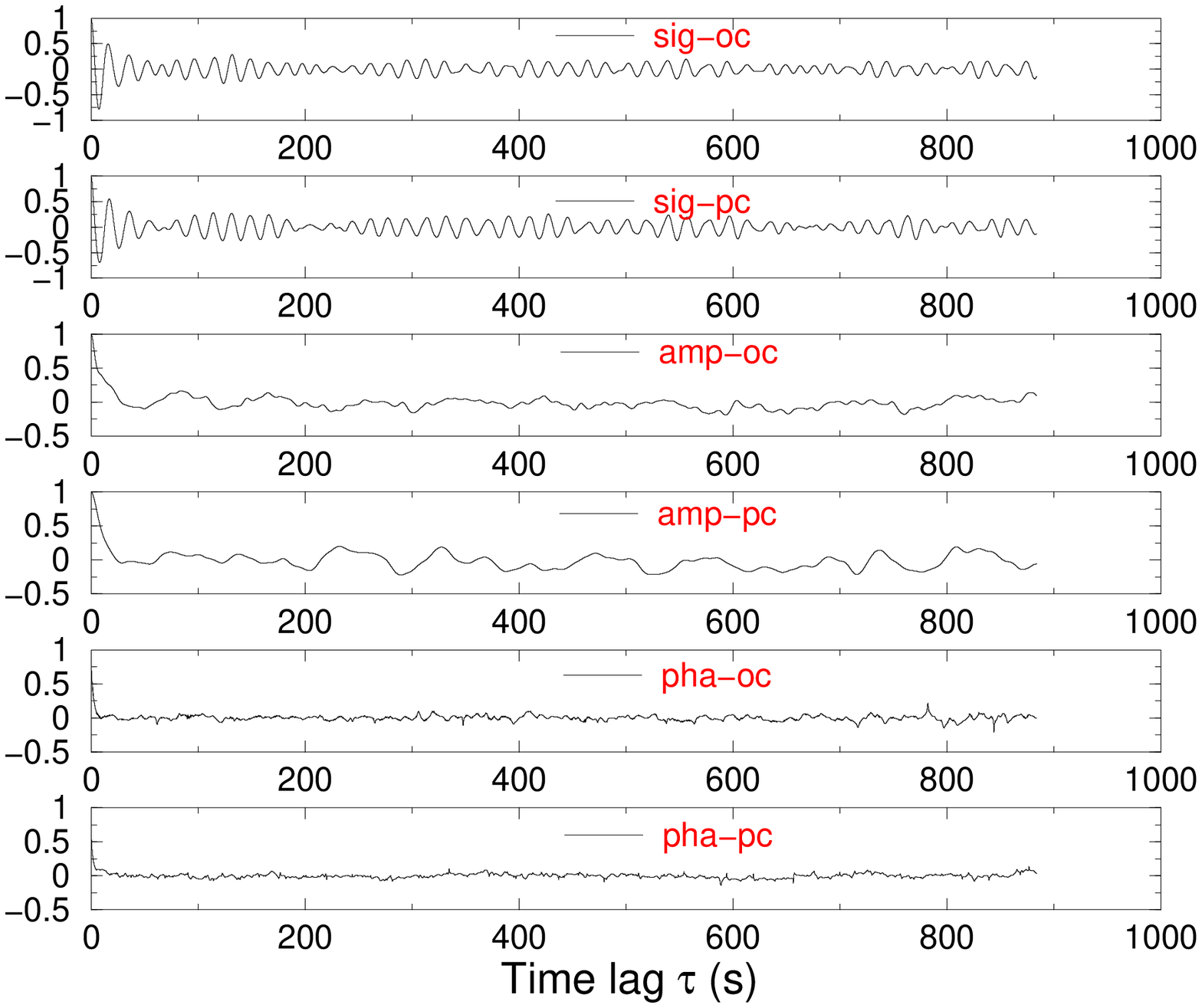}}}
\epsfysize=0.8\columnwidth{{\epsfbox{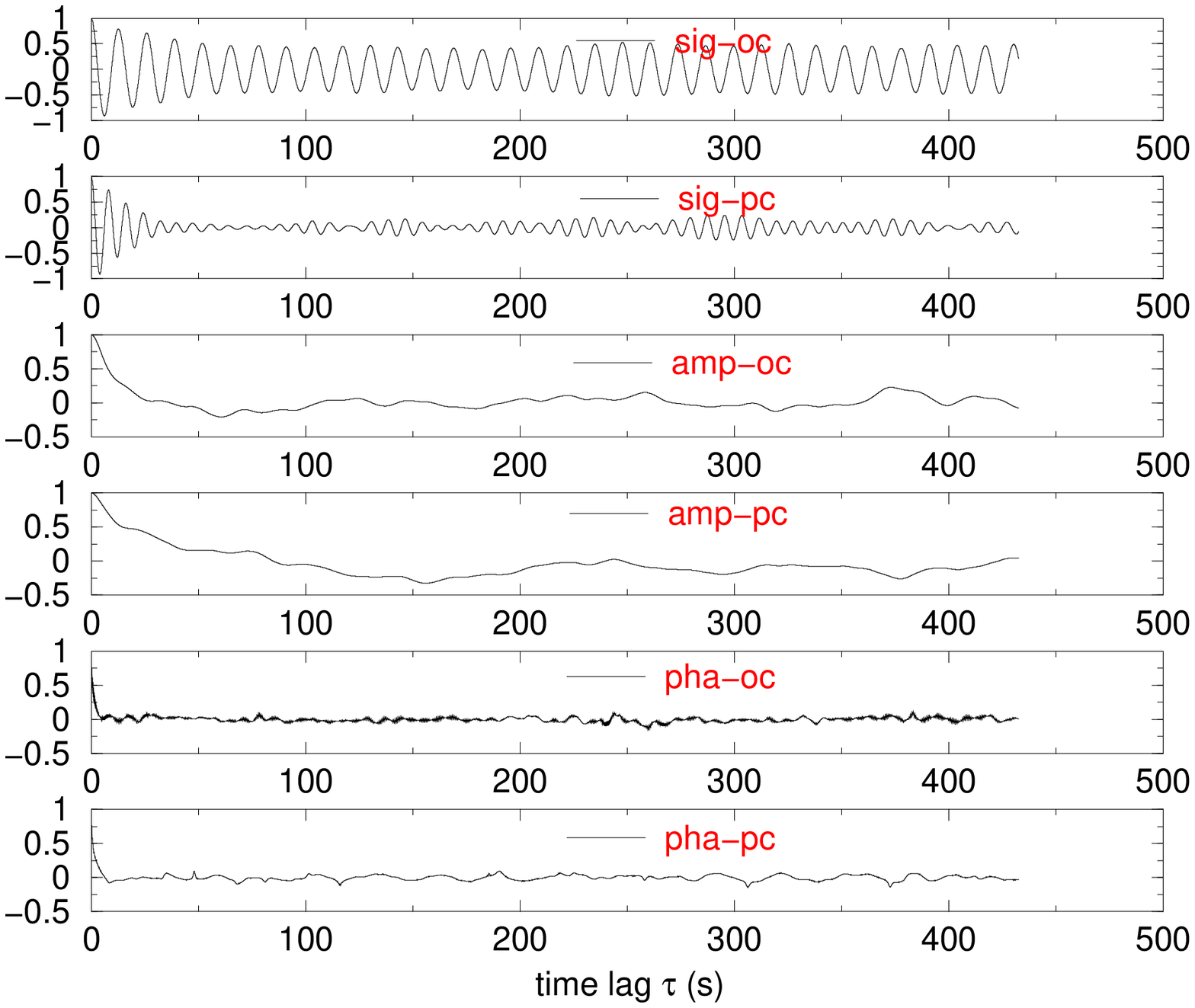}}}
\caption{Auto-correlation functions (y axis) $C(\tau)$ for the original
  data and their instantaneous amplitudes and phase increments for the
  group JL839 (up) and the group GX1129 (down). }
\label{fig4}
\end{figure}

Here we investigate the auto-correlation functions for the original
data and their instantaneous amplitudes and phases in OC and PC
directions. The results in most groups can be summarized in the upper
panel of Fig.~\ref{fig4}, where we mark the results for the original
signal as ``sig'', the results for the instantaneous amplitude as
``amp'', and the results for the instantaneous phase increments as
``pha''. Note that in conventional methods only the correlation of the
original signal was considered.

As shown in the upper panel of Fig.~\ref{fig4}, for most groups of
data we find that the periodicity is presented in the auto-correlation
functions $C(\tau)$ of the original signal. The period shown in the
correlation function is approximately the averaged period of the
original signal. Thus when the time lag $\tau$ increase from zero,
$C(\tau)$ will decrease rapidly from 1 to a small value. Different
from that of the original signal, auto-correlations of both
corresponding instantaneous amplitude and phase increment do not show
apparent periodicity as a function of the time lag $\tau$. $C(\tau)$
will normally decrease as the time lag $\tau$ increases, and finally
$C(\tau)$ would fluctuate around zero. Very interestingly the
correlation length (in unit of measurement time) for the instantaneous
amplitude is around one period of the original signal. However the
situation is different for the instantaneous phase increment. As shown
in Fig.~\ref{fig4}, such correlation length is only 1/10 of one period
of the original signal. Thus the correlation in the instantaneous
phase increment may be lost for even two patches in the original
signal which are in the same period in time and are not very far
away. In contrast, the correlation in the instantaneous amplitude is
usually kept for two patches in the original signal which are in the
same period.

It is very striking that the data in the group GX1129 display two
different properties from those shown above. We find that in
horizontal (OC) direction the original signal presents very strong
correlations even when the time lag $\tau$ is huge. In vertical (PC)
direction, the correlation length of the instantaneous amplitude
series becomes much larger than one period of the original
signal. These two results both imply that the ship is not interrupted
by any considerable actions when the measurement is processing, thus
the characteristics of the system could be maintained in time. This
clarification is verified through the conversations with our
collaborators: this group of data are actually obtained when the ship
navigates freely.
 
%\begin{figure}
%\epsfysize=0.6\columnwidth{{\epsfbox{self-correlation_GX1129.eps}}}
%\caption{Autocorrelation functions (y axis) for the original data and their
%  instantaneous amplitudes and phase increments of the group GX1129. }
%\label{fig5}
%\end{figure}

\subsection{Cross-correlations for the original data and their instantaneous amplitudes and phase increments}\label{crosscorr}

\begin{figure}
\epsfysize=0.8\columnwidth{{\epsfbox{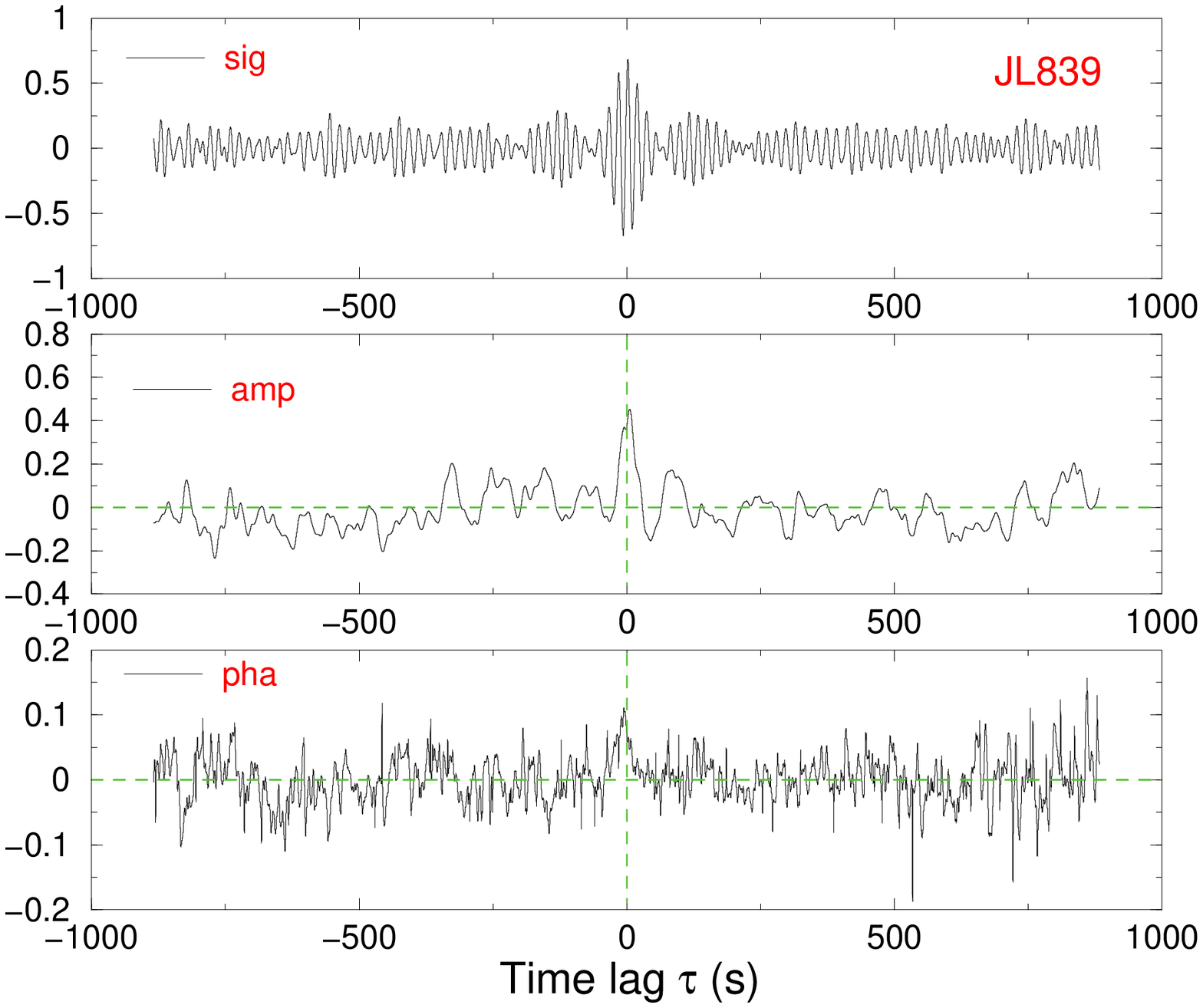}}}
\epsfysize=0.8\columnwidth{{\epsfbox{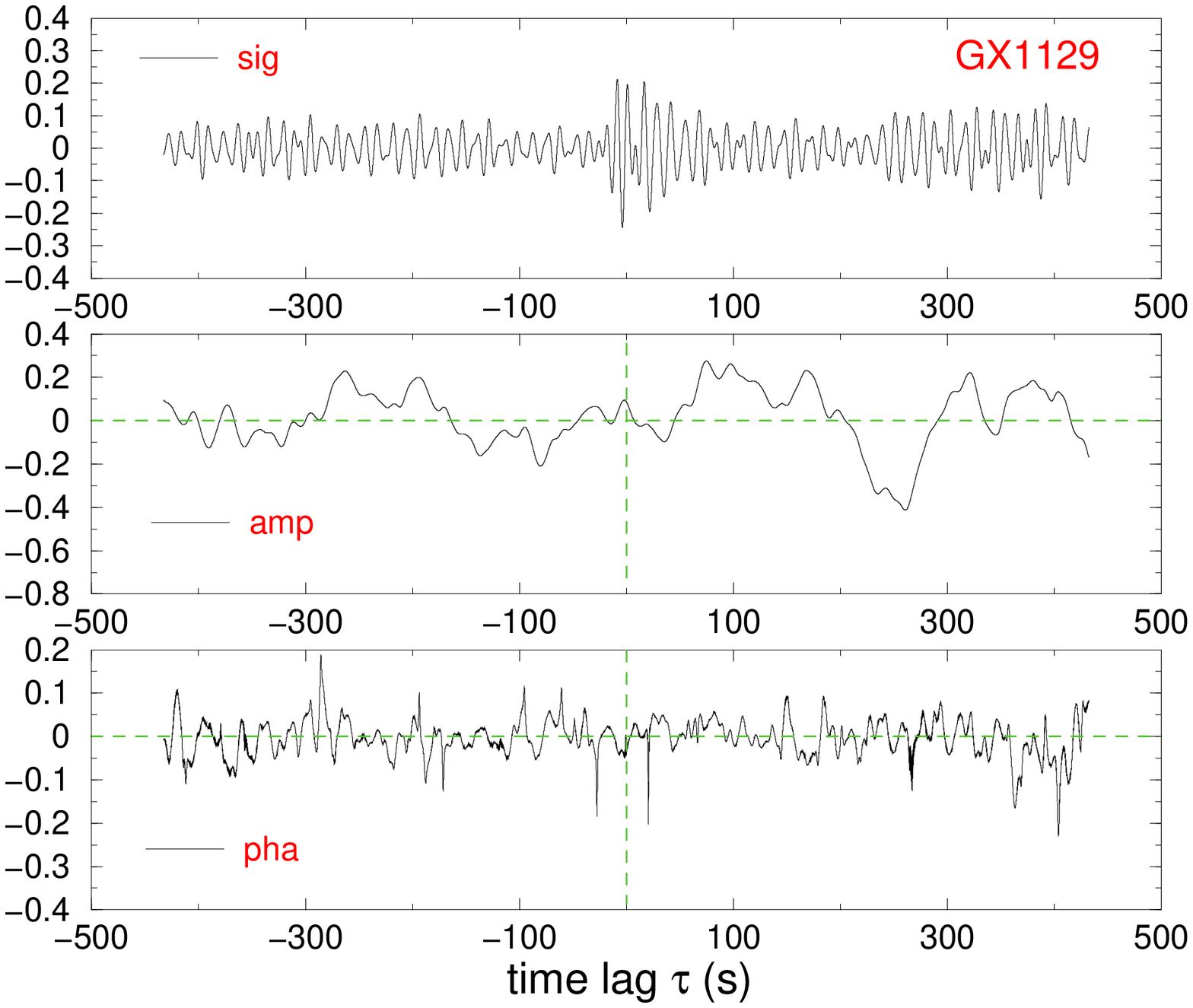}}}
\caption{The OC-PC cross-correlation functions (y axis) for the original
  data and their instantaneous amplitudes and phase increments for the
  groups JL839 (up) and GX1129 (down). }
\label{fig5}
\end{figure}

Similar to those we present in the last section, the
cross-correlations for the original data and their instantaneous
amplitudes and phase increments also present some distinct
properties. The typical case for most groups is shown in the upper
panel of Fig.~\ref{fig5}. The OC-PC cross-correlation is calculated
utilizing the following formula:
\begin{eqnarray}
C_{\mbox{cross}}(\tau) = \langle(s_{OC}(t)-\langle s_{OC} \rangle) 
  (s_{PC}(t+\tau)-\langle s_{PC} \rangle) \rangle,
\label{eqn3}
\end{eqnarray}
where $\tau$ is the time lag.

As shown in Fig.~\ref{fig5}, the group GX1129 continue presenting its
special properties in cross-correlations of signals in different
directions. For the cross-correlations of the original signals, it is
apparent that $C_{\mbox{cross}}(\tau)$ behave differently for positive
and negative time lag $\tau$. For positive $\tau$ we observe much
stronger correlations, which implies that the ship-rocking in the
horizontal (OC) direction at the present time will affect more
effectively to the ship-rocking in the vertical (PC) direction at the
future time. Thus the ship-rocking in the horizontal direction is the
one which provides the ``power'' to ship-rocking in the vertical
direction. Such result explains why in the last section the increase
in the correlation length of the instantaneous amplitudes happens in
the vertical direction when the ship navigates freely, while in the
horizontal direction the correlation length is almost unchanged.

In contrast, for other groups we find that $C_{\mbox{cross}}(\tau)$ is
symmetric for positive and negative small time lag $\tau$, as shown in
Fig.~\ref{fig5}. Thus there is no priority in vertical or horizontal
direction for the ship-rocking. As a result at $\tau=0$ we find that
$C_{\mbox{cross}}(\tau)$ achieves absolute maximum for the
instantaneous amplitudes. However for the group GX1129, at $\tau=0$
the cross-correlation $C_{\mbox{cross}}(\tau)$ is close to zero. The
conclusion is similar for the cross-correlation of the instantaneous
phase increments, though when $\tau=0$ we observe that
$C_{\mbox{cross}}(\tau)$ only achieves local maximum.

\section{Prediction for the ship-rocking at the future time}\label{predict}

In reality, the ability of the prediction to the ship-rocking signal
at the future time is very crucial in the ocean-based flight vehicle
measurement. However, with the accuracy of 20 arc seconds, the current
available methods based on the stochastic processes could predict only
the behavior of the ship-rocking signal in the very near future, i.e.,
less than 200 ms~\cite{ARMA}. In addition, the possibility to improve
this kind of methods is very limited since such methods contain many
adjusting parameters. In Sec.~\ref{hilbert}, we find that the
evolution of the analytic signal $S(t)$ is very slow when seen in the
complex plane. Based on this fact and results in Sec.~\ref{amp_pha},
here we set up a new method to make predictions to behaviors of
the ship-rocking signal at certain future times.

\begin{figure}
\epsfysize=0.9\columnwidth{\rotatebox{-90}{\epsfbox{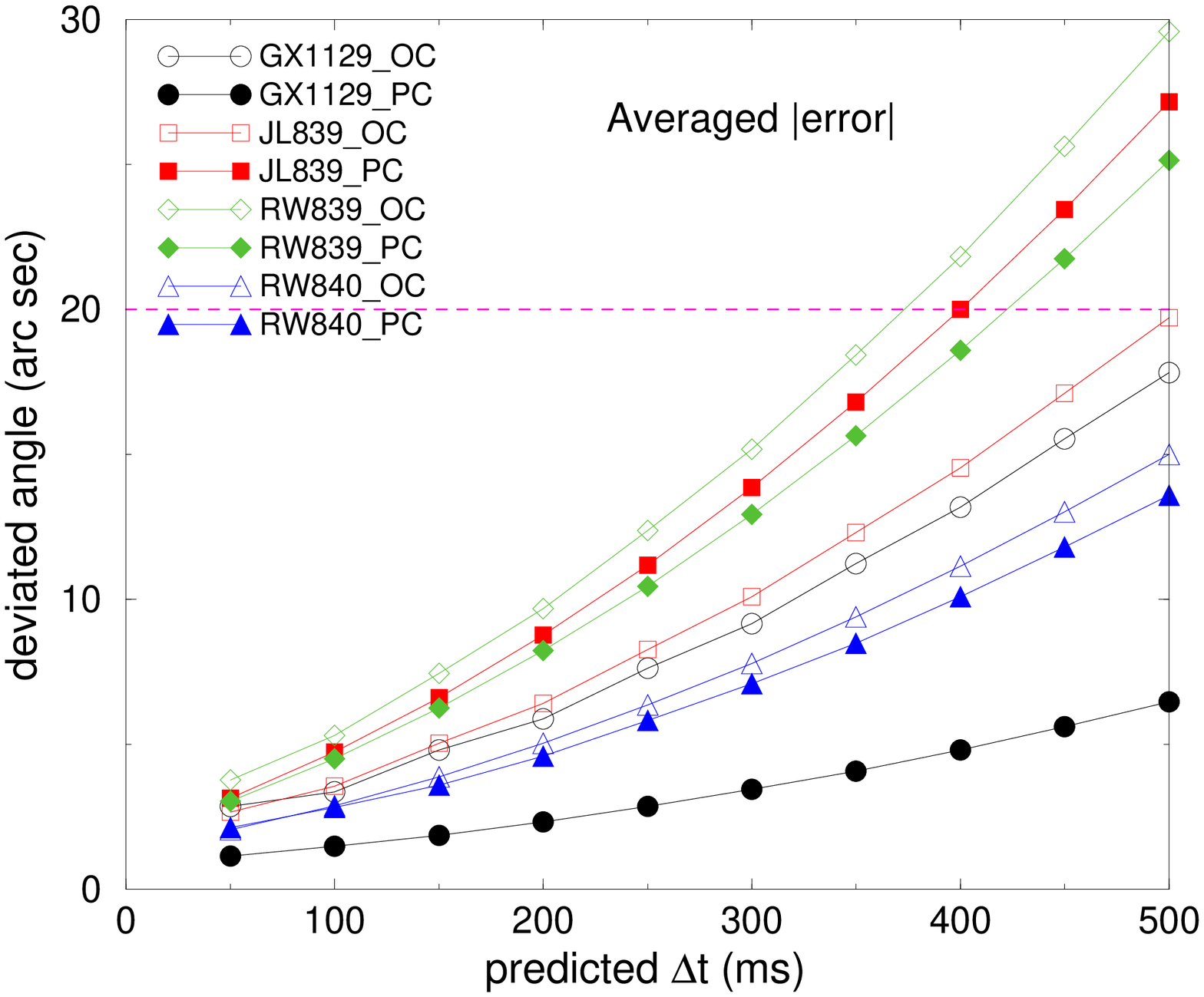}}}
\epsfysize=0.9\columnwidth{\rotatebox{-90}{\epsfbox{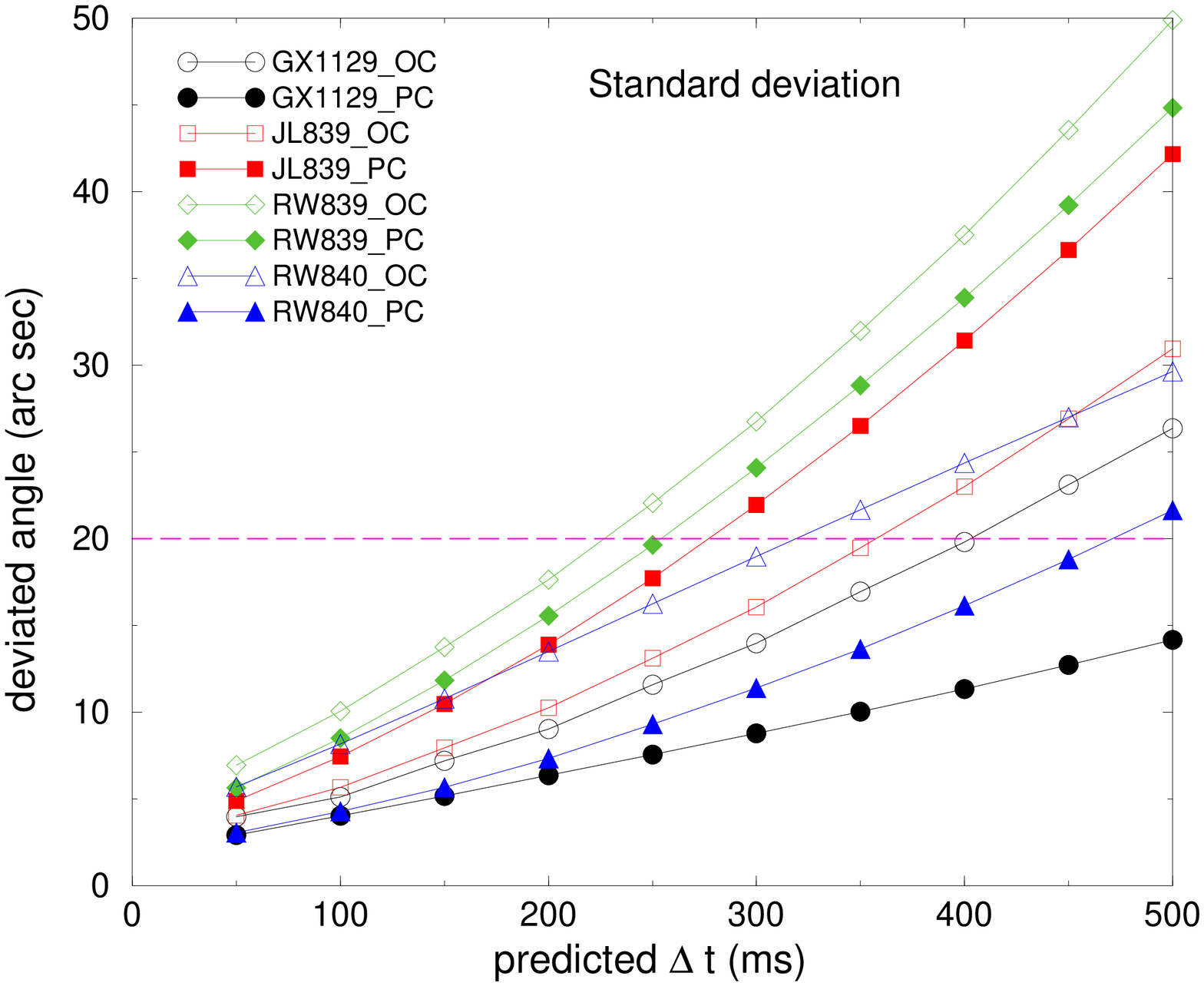}}}
\caption{The absolute errors and the standard deviations of the
  prediction from the Hilbert transform based method as a function of
  the prediction time interval. }
\label{fig6}
\end{figure}

In an initial setup we provide here a simple model. In two-dimensional
plane one can use two polar-coordinate parameters $\rho$ and $\theta$
to quantify the behavior of a time series. For the data of interest,
$\rho$ corresponds to the instantaneous amplitude, and $\theta$
corresponds to the instantaneous phase. From Sec.~\ref{hilbert} and
Sec.~\ref{amp_pha}, we have learned that the analytic signal usually
evolve very smoothly in time on the complex plane. On the other side,
the ability of our prediction is also limited by the correlation in
signals. From Sec.~\ref{autocorr} and Sec.~\ref{crosscorr}, we
estimate that the maximal time range in which we could obtain
convincing prediction is around 1/10 of the periods in the original
signal. In such a short time range we utilize the simplest model in
the polar coordinate: the helix model. In this model $\rho$ evolve
linearly with the change in $\theta$. Based on the results in
Sec.~\ref{amp_pha} we assume that the instantaneous phase also evolves
linearly in time. In reality this assumption is invalid only at the
points where we observe outliers, as shown in in Fig.~\ref{fig3}.

From this model we propose to utilize 10 known values in the original
time series to predict values of the original time series at one to
ten following measurement time. (Note that the data are measured every
50 milliseconds). The data measured before these ten known values are
also considered known values. From these known values we first
construct its corresponding analytic signal. We next extend the track
of the analytic signal on the complex plane from the helix model. The
ten predicted values are then compared to the actual values in the
original time series. The results of comparison are shown in
Fig.~\ref{fig6}. From the results of both the averaged absolute error
and the standard deviation we find that our predicted values are very
close to that of the original signals within a much wider range of the
prediction time window than that of current methods.  The results
seem better in the averaged absolute error than that in the standard
deviation. We attribute this to certain outliers in the instantaneous
phase increments. Some tunning at positions near the outliers should
be able to further improve the current accuracy.

\section{Summary}

\noindent In summary, we have analyzed four groups of ship-rocking
signals in horizontal and vertical directions utilizing a Hilbert based
method from statistical physics. Our method gives a way to construct
an analytic signal on the two-dimensional plane from a time
series. The analytic signal share the complete property of the
original time series. From the analytic signal of a time series, we
have found some information of the original time series which are
often hidden from the view of the conventional methods. For the
current data, their analytic signals usually evolve very smoothly on
the complex plane. In addition, the phase of the analytic signal is
usually moves linearly in time. From the auto-correlation and
cross-correlation functions of the original signals as well as the
instantaneous amplitudes and phase increments of the analytic signals
we have found that the ship-rocking in horizontal direction drives the
ship-rocking in vertical direction when the ship navigates freely. And
when the ship keeps a fixed navigation direction such relation
disappears. Based on these results we could predict certain amount of
future values of the ship-rocking time series based on the current and
the previous values. Our predictions are as accurate as the
conventional methods from stochastic processes and provide a much
wider prediction time range.

This work is supported by the JLOMAC fund NO.FOM2014OF002, the
National Natural Science Foundation of China (Grants no. 11275184 and
61403421), Anhui Provincial Natural Science Foundation (Grant
no. 1208085MA03), and the Fundamental Research Funds for the Central
Universities of China (Grants no. WK2030040012 and 2340000034).


\begin{thebibliography}{99}
\bibitem{ChenPRE06} Z. Chen, {\it et al.} {\it Phys. Rev. E} {\bf 73},
  031915 (2006).

\bibitem{PengPRE94} C.-K. Peng, {\it et al.} {\it Phys. Rev. E} {\bf
  49}, 1685-1689 (1994).

\bibitem{ChenPRE02} Z. Chen, P.Ch. Ivanov, K. Hu, and H.E. Stanley. {\it Phys. Rev. E} {\bf 65}, 041107 (2002).

\bibitem{ChenPRE05} Z. Chen, {\it et al.} {\it Phys. Rev. E} {\bf 71}, 011104 (2005).

\bibitem{ARMA} Xiaoyong Li, Zhijun Du and Guiming Chen. {\it Manned
  Spaceflight} {\bf 3}, 38-43 (2005). (in Chinese)

\end{thebibliography}
\end{document}